# Strongly Enhanced Raman Optical Activity in Molecules by Magnetic Response of Nanoparticles

*Tong Wu[1], Xiuhui Zhang[2], Rongyao Wang[1] and Xiangdong Zhang[1]\**

[1]School of Physics and Beijing Key Laboratory of Nanophotonics & Ultrafine Optoelectronic Systems, Beijing Institute of Technology, Beijing, 100081, China

[2]School of Chemistry and Key Laboratory of Cluster Science of Ministry of Education, Beijing Institute of Technology, Beijing, 100081, China

**ABSTRACT:** An analytical theory for the surface-enhanced Raman optical activity (SEROA) with the magnetic response of the substrate particle has been presented. We have demonstrated that the SEROA signal is proportional to the magnetic polarizability of the substrate particle, which can be significantly enhanced due to the existence of the magnetic response. At the same time, a large circular intensity difference (CID) for the SEROA can also be achieved in the presence of the magnetic response. Taking Si nanoparticles as examples, we have found that the CID enhanced by a Si nanoparticle is 10 times larger than that of Au. Furthermore, when the molecule is located in the hotspot of a Si dimer, CID can be 60 times larger. The phenomena originate from large magnetic fields concentrated near the nanoparticle and boosted magnetic dipole emission of the molecule. The symmetric breaking of the electric fields caused by the magnetic dipole response of the nanoparticle also plays an important role. Our findings provide a new way to tailor the

Raman optical activity by designing metamaterials with the strong magnetic response.

**I.INTRODUCTION**

Chirality plays a crucial role in modern biochemistry and the evolution of life.[1] Many biologically active molecules are chiral, detection and quantification of chiral enantiomers of these biomolecules are of considerable importance. In the past years, many spectroscopic techniques have been proposed for the determination of the molecule chirality, including electronic circular dichroism (ECD), vibrational circular dichroism (VCD), and Raman optical activity spectroscopy (ROA).[2-4] Among all these techniques, Raman optical activity spectroscopy (ROA) is a powerful method,[5] because this technique can give the chirality related to the structural information of all parts of the molecule and be particularly sensitive to the conformation and dynamics of biological molecules.[6,7] For example, the ROA has been proved to be useful in characterizing secondary order structures of proteins,[8] determining the absolute configurations of small chiral molecules,[9] and studying the dynamics of biomolecules.[10]

However, the widely use of the ROA technique is hampered by the weakness of signal intensities, which is always 3 orders of magnitude smaller than its parent Raman intensities.[4-11] Usually long measuring times and densely concentrated samples are required to guarantee the reliability of the measurement. How to improve the detection efficiency with the ROA technique becomes a key problem in recent years. Many studies focused on the surface enhanced Raman optical activity (SEROA) based on metal surface plasmon resonances.[11-30] This is because the Raman scattering from molecules placed near metal surfaces can be strongly enhanced, giving rise to surface-enhanced Raman scattering (SERS).[31] The typical enhancement factor of the SERS is about $10^7$ and can reach $10^{14}$ under favorable conditions.[31,32] This large enhancement can be understood to be dominated by electromagnetic mechanism (EM) due to resonances of the incident beam with metal surface plasmon excitations.

The first SEROA theory was purposed by Efrima who showed that in addition to enhancements caused by large evanescent electric fields at the metal surface, electric field gradient also plays a key role in amplifying the ROA signal.[21,22] Subsequently, Janesko and Scuseria considered the effect of multipole responses of substrates on the SEROA.[23] They found that electric field gradient contributions might be larger for particles with the quadrupolar response. At the same time, the ROA signal is also very sensitive to the orientation of the molecule-substrate.[28,29] A recent work by Chulhai and Jensen,[28] showed that, for chiral molecules with fixed orientations respecting with the surface of the nanoparticle (NP), the field gradient causes significant change in the SEROA spectrum, which prevents the observation of mirror-imagine SEROA in the real experiment. This problem may be settled by using systems with random molecule-mental orientations.[33] However, according to the previous theory,[25,28,29] the SEROA signals from different molecules tend to cancel in an ensemble measurement, and the measured circular intensity difference (CID) is supposed to be much smaller than that of the pure molecule.

Over the past years, it has been long believed that the magnetic component is not expected to make a large contribution to the SEROA.[11,21,22] Furthermore it is also difficult to obtain strong magnetic response at visible and near-infrared frequencies because the magnetic susceptibility of all natural materials tails off at microwave frequencies. The magnetic response of the NP is often neglected in most developed SEROA theories. Nevertheless, recent investigations have shown that the magnetic response at visible frequencies can be achieved by designing metamaterials. For example, specially designed metallic nanostructures, such as Metal−Insulator−Metal (MIM) dimers,[34] plasmonic nanorings or nano cups, can exhibit magnetic dipole responses at the optical frequency.[35,36] It is interesting that Si NPs have been proved, both theoretically and experimentally,[37,38] to possess a strong magnetic dipolar response originating from circular displacement currents driven by an incident electric field in the optical frequency.[38] Moreover, recent works have also shown that the magnetic field plays a pivotal role in the probing of the circular dichroism (CD) of chiral molecules.[39-41]

Motivated by above investigations, in this work we discuss the effect of magnetic responses on the SEROA. An analytical theory for the SEROA with the magnetic response of the substrate particle has been presented. We find that the SEROA signal can be significantly enhanced due to the existence of the magnetic response, and a large CID can be achieved. Taking Si NPs as model systems, we have demonstrated that CID for the SEROA can be enhanced by 10 times for a Si NP compared with the corresponding Au NP, and more than 60 times enhancement can be achieved when the molecule is located in the hotspots of a Si dimer. The physical origins for these phenomena have been discussed.

## II. ANALYTICAL THEORY OF SEROA BY A NANOPARTICLE WITH THE MAGNETIC RESPONSE

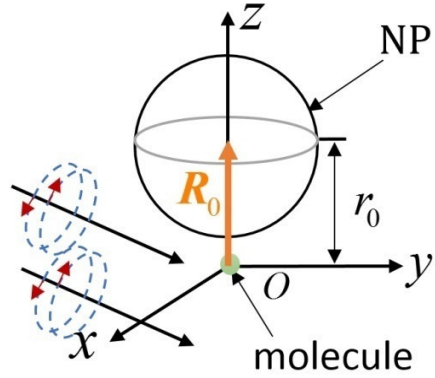

**Figure 1.** Geometry and coordinate of a hybrid system consisting of a chiral molecule and nanoparticle (NP). The chiral molecule is put at the origin with electric dipole ($\boldsymbol{\mu}$), magnetic dipole ($\boldsymbol{m}$) and traceless electric quadrupole ($\hat{\theta}$). The NP with radius of $R_s$ is located in $\boldsymbol{R}_0 = (0,0,r_0)$ with induced electric dipole ($\boldsymbol{\mu}^s$) and magnetic dipole ($\boldsymbol{m}^s$).

We consider a hybrid system consisting of a chiral molecule and a NP as shown in Figure 1, which is excited by circularly polarized plane waves with the angular frequency $\omega_0$. The chiral molecule is put at the origin of the coordinate, the spherical NP locates at $\boldsymbol{R}_0 = (0,0,r_0)$ with radius $R_s$, which is much smaller than the wavelength of the excitation wave. The molecule used here is assumed as a combination of electric dipole ($\boldsymbol{\mu}$), magnetic dipole ($\boldsymbol{m}$) and traceless electric quadrupole ($\hat{\theta}$). Within off-resonance condition and semi-classical approximation, the corresponding elements for $\boldsymbol{\mu}$, $\boldsymbol{m}$ and $\hat{\theta}$ are given by

$$\mu_a \approx \tilde{\alpha}_{ab} E_b^I + \tilde{G}_{ab} B_b^I + \frac{1}{3} \tilde{A}_{a,bc} \partial_b E_c^I \tag{1}$$

$$m_a \approx -\tilde{G}_{ba} E_b^I \tag{2}$$

$$\theta_{ab} \approx \tilde{A}_{c,ab} E_c^I \tag{3}$$

in the Cartesian coordinate.[2-5] The subscripts $a$, $b$ and $c$ represent the coordinate components $x$, $y$ or $z$, respectively. Einstein summation convention and MKS system of units are employed throughout the article. Here $E_b^I$ and $B_b^I$ are the total electric and magnetic fields at the position of the molecule. The $\tilde{\alpha}_{ab}$, $\tilde{G}_{ab(ba)}$ and $\tilde{A}_{a(b),bc(cb)}$ represent elements of dynamic molecular response tensors, which can be calculated by using time-dependent perturbation theory and Placzek approximation.[42] For concision, response tensors are written here instead of their derivations with respect to normal coordinates. Here we presume that the radius $R_s$ of the NP is not extremely small ($R_s$>5nm), so that higher order terms, like the electric quadrupole−quadrupole contribution, can be safely neglected.[28] The subwavelength NP can be modeled as a point, then, the isotropic and frequency-dependent electric and magnetic responses are expressed as

$$\mu_a^s(\omega) = \alpha_e^s(\omega) E_a(\mathbf{R}_0) \tag{4}$$

$$m_a^s(\omega) = \alpha_m^s(\omega) B_a(\mathbf{R}_0) \tag{5}$$

where $\omega$ is the frequency, $\alpha_e^s = -6i \frac{\varepsilon_0 \varepsilon}{k^3} \pi T_1^{(E)}$ and $\alpha_m^s = -6 \frac{1}{k^3} \pi i T_1^{(M)}$ are scalar electric and magnetic polarizabilities, $\varepsilon$ and $\varepsilon_0$ are the relative and vacuum permittivity, respectively, and $k$ is the wave vector in the medium. The $T_1^{(E)}$ and $T_1^{(M)}$ are elements of the Mie scattering matrix of the spherical particle, which have been given in ref 43. This is in contrast to previous theories, where only electric dipole and electric quadrupole responses are taken into account.[21,23,24,25,28] The EM fields at the position of the molecule are strongly mediated by these responses of the NP, which can be written as:

$$E_a^I = E_a^i + \frac{\mu\mu_0}{4\pi} e^{ikr_0}\left(\mu_a^s(\omega_0)C_1(\omega_0) - \frac{n_a n_b}{n^2}\mu_b^s(\omega_0)C_2(\omega_0) - \frac{\varepsilon_{abc}n_b}{c}m_c^s(\omega_0)D_1(\omega_0)\right) \quad (6)$$

$$B_a^I = B_a^i + \frac{\mu\mu_0}{4\pi} e^{ikr_0}\left(m_a^s(\omega_0)\frac{n^2}{c^2}C_1(\omega_0) - n_a n_b m_b^s(\omega_0)\frac{1}{c^2}C_2(\omega_0) + \frac{\varepsilon_{abc}n_b}{c}\mu_c^s(\omega_0)D_1(\omega_0)\right) \quad (7)$$

where $E_a^i$ and $B_a^i$ represent components of electric and magnetic fields of the incident wave at the origin, $c$ is the velocity of light in vacuum, $n$ is the refraction index of the embedded medium, $\mu$ is the relative permeability in the medium, and $n_{a(b)} = nR_{0\,a(b)}/r_0$. The induced electric and magnetic dipole moments on the NP can be calculated with $\mu_{a(bc)}^s(\omega_0) = \alpha_e^s(\omega_0)E_{a(bc)}^i\exp(i\mathbf{k}\cdot\mathbf{R}_0)$ and $m_{a(bc)}^s(\omega_0) = \alpha_m^s(\omega_0)B_{a(bc)}^i\exp(i\mathbf{k}\cdot\mathbf{R}_0)$, where $\mathbf{k}$ is the wave vector of the incidental plane wave. The last three terms in eqs 6 and 7 are the scattering fields of the NP. Expressions of $C_1(\omega_0)$, $C_2(\omega_0)$ and $D_1(\omega_0)$ are given by

$$C_1(\omega_0) = \left(\frac{\omega_0^2}{r_0} + \frac{i\omega_0 c}{nr_0^2} - \frac{c^2}{n^2 r_0^3}\right) \quad (8)$$

$$C_2(\omega_0) = \left(\frac{\omega_0^2}{r_0} + \frac{3i\omega_0 c}{nr_0^2} - \frac{3c^2}{n^2 r_0^3}\right) \quad (9)$$

$$D_1(\omega_0) = \left(\frac{\omega_0^2}{r_0} + \frac{i\omega_0 c}{nr_0^2}\right) \quad (10)$$

It's worth to note that, different from previous works,[21-25,28] which are based on quasi-static approximation and fields proportion to $1/r_0^3$ are only considered, the electric and magnetic fields are written in all wave zones including the near-field, intermediate, and far-field regions.[44] Later we will see the electric field in the intermediate zone plays a crucial role in enhancing the ROA signal.

The radiation field at the Raman shifted frequency $\omega_p$, originated from the multipole excitation of the molecule, can induce dipole responses on the spherical NP according to eqs 4 and 5. Thus, the total multipoles of the molecule-NP hybrid system are given by

$$\mu_a^D = \mu_a + \mu_a^s(\omega_p) \tag{11}$$

$$m_a^D = m_a + m_a^s(\omega_p) - \frac{1}{2}i\omega_p \varepsilon_{abc} R_{0b} \mu_c^s(\omega_p) \tag{12}$$

$$\theta_{ab}^D = \theta_{ab} + \frac{3}{2}\mu_a^s(\omega_p) R_{0b} + \frac{3}{2}\mu_b^s(\omega_p) R_{0a} - \mu_c^s(\omega_p) R_{0c} \delta_{ab} \tag{13}$$

where $\varepsilon_{abc}$ is the Levi-Civita tensor, $\delta_{ab}$ is the Kronecker delta function. Here the expansions of multipoles in eqs 11-13 have been realized with respect to the position of the molecule (origin of the coordinate as shown in Figure 1). The third term in eq 12 and last three terms in eq 13 are caused by phase differences between the fields scattered by the molecule and NP multiple moments. They are added to achieve the origin-independence of our theory.

Inserting eqs 4-5 into eqs 11-13 and using the expressions for the multipole radiation fields to get $\mu_a^s(\omega_p)$ and $m_a^s(\omega_p)$, the total multipoles can be evaluated. The concrete expressions are provided in the Supporting Information 1.

Based on the above analysis, we calculate the ROA intensity for the incident circular polarization experimental setup (SEICP-ROA), which is given by[2-5,28]

$$I^{(p)}_{SEICP-ROA} = K_p K_r \sum_{D=R;L} (-1)^\kappa \left| E_{rad,t,a}(D) \right|^2 \tag{14}$$

where $D$ denotes whether the incident wave is right ($R$) or left ($L$) handed, $\kappa = 0$ when $D = R$, $\kappa = 1$ when $D = L$, $K_r = 8\left|\pi/(\omega_p^2 \mu_0)\right|^2$ with $\mu_0$ being the vacuum permeability, $k$ is the wave vector in the embedded medium and $R_d$ is the distance between the detector and the origin of the coordinate. $K_p$ is a normalizing factor which is defined as[30,31]

$$K_p = \frac{\pi^2}{\varepsilon_0^2}(v_0 - v_p)^4 \frac{h}{8\pi^2 c v_p} \frac{1}{1-\exp\left[-hcv_p/k_B T\right]} R_d^2 \tag{15}$$

where $v_0$ and $v_p$ are the absolute wave-number of the incident light and the $p^{th}$ vibrational mode, respectively, $T$ is temperature (298K), $\varepsilon_0$, $c$, $k_B$ and $h$ are the universal physical constants. In eq15, $E_{rad,t,a}$ is the Raman scattering electric field, which is given by

$$E_{rad,t,a}(D) \simeq E^{(1)}_{rad,t,a}(\tilde{\alpha},D) + E^{(2)}_{rad,t,a}(\tilde{G},D) + E^{(2)}_{rad,t,a}(\tilde{A},D) \tag{16}$$

where $E^{(1)}_{rad,t,a}(\tilde{\alpha},D)$ is a linear function of $\tilde{\alpha}$, while $E^{(2)}_{rad,t,a}(\tilde{G},D)$ and $E^{(2)}_{rad,t,a}(\tilde{A},D)$ relate only with dynamic molecular response tensors $\tilde{G}$ and $\tilde{A}$, respectively. Their expressions are very complex, which are given in Supporting Information 2. In fact, $I^{(p)}_{SEICP-ROA}$ comes from three aspects of contributions: $\tilde{\alpha}\tilde{\alpha}$ interaction $I^{(p)-\alpha\alpha}_{SEICP-ROA}$, $\tilde{\alpha}\tilde{G}$ interaction $I^{(p)-\alpha G}_{SEICP-ROA}$ and $\tilde{\alpha}\tilde{A}$ interaction $I^{(p)-\alpha A}_{SEICP-ROA}$, which are determined by the products of molecular response tensors. Usually in order to guarantee the chiral selectivity of the ROA spectrum, $I^{(p)-\alpha\alpha}_{SEICP-ROA}$ should vanish in the ROA signal, which can be accomplished by taking orientation averages over directions of incident waves and detectors.[26,27] Resonant or near resonant molecules with asymmetric $\tilde{\alpha}$ response tensors are exceptions,[45] which are beyond the discussion of our work. Thus,

$$\bar{I}_{SEICP-ROA}^{(p)} = \bar{I}_{SEICP-ROA}^{(p)-\alpha G} + \bar{I}_{SEICP-ROA}^{(p)-\alpha A} \tag{17}$$

where

$$\bar{I}_{SEICP-ROA}^{(p)-\alpha \mathcal{M}} = \left\langle K_p K_r \sum_{D=R;L} (-1)^{\kappa} \left[ E_{rad,t,a}^{(1)}(\tilde{\alpha},D)\left(E_{rad,t,a}^{(2)}(\mathcal{M},D)\right)^* + E_{rad,t,a}^{(2)}(\mathcal{M},D)\left(E_{rad,t,a}^{(1)}(\tilde{\alpha},D)\right)^* \right] \right\rangle_{\Omega} \tag{18}$$

Here $\mathcal{M}$ represents either $\tilde{G}$ or $\tilde{A}$, and the bar above '$I$' indicates that the average is taken over the solid angle of the incident field as denoted by '$\langle \cdots \rangle_{\Omega}$' on the right side of the equation. $E_{rad,t,a}^{(1)}(\tilde{\alpha},D)$, $E_{rad,t,a}^{(2)}(\tilde{G},D)$ and $E_{rad,t,a}^{(2)}(\tilde{A},D)$ should be evaluated at the position of the detector which is put at infinity being opposite with the propagating direction of the incident wave.

The expressions of $E_{rad,t,a}^{(1)}(\tilde{\alpha},D)$, $E_{rad,t,a}^{(2)}(\tilde{G},D)$ and $E_{rad,t,a}^{(2)}(\tilde{A},D)$ include many terms, which are given in the Supporting Information 2. We systematically analyzed the orders of their magnitudes (see Supporting Information 3 for more details of analyses). This has been done by, firstly, substituting the expression of each term with some simple variables which have the same order of magnitude. Later, these terms are classified by their orders of magnitudes. Terms with the leading order are $\lambda / R_s$ times larger than those with the second leading order, and the sequence goes on. Since the multiplication between two leading order terms are zero due to the orientation averaging of the incident waves,[23] the product from the multiplication between the leading order term and the second leading order terms become the maximum in these expressions, they are far greater than any other products. Thus, we take these terms and the SEROA scales as

$$\overline{I}_{SEICP-ROA}^{(p)-\alpha G} = \mathscr{O}\left[\overline{I}_{SEICP-ROA}^{(p)-\alpha A}\right] = K_p \mathscr{O}\left[kd_0 S_e^{\ 3}(1+S_m)\tilde{\alpha}^2\right] \tag{19}$$

where $S_e = \dfrac{\mu\mu_0 c^2}{4\pi r_0^3}\alpha_e^s$ and $S_m = \dfrac{\mu\mu_0}{4\pi r_0^3}\alpha_m^s$ are dimensionless quality which are proportional to the electric and magnetic dipole responses of the NP. The $d_0$ denotes a length on the order of molecular dimensions. In deriving eq 19, the retardation of the plane wave and the phase differences between the fields scattered by the NP and molecule are ignored. A discussion on the influence of these effects is provided in the Supporting Information 4. The term including $S_m$ in eq 19 indicates that the SEROA is strongly enhanced by the magnetic dipole response of the NP. If $S_m$ is larger than 1, the SEROA signal is supposed to be much stronger than that enhanced by a traditional plasmonic NP with only the electric dipole response. Our theory, thus, predicts that the ROA signal can be significantly amplified by nano structures with the strong magnetic response, like split ring, nano bowl or Raspberry-like metamolecule.[36,46,47] This is one of main contributions of this work.

Here it's worth noting that even though both $\overline{I}_{SEICP-ROA}^{(p)-\alpha G}$ and $\overline{I}_{SEICP-ROA}^{(p)-\alpha A}$ are proportional to the magnetic dipole response of the NP, different mechanisms should be assigned in interpreting origins of the phenomenon. The enhancement of $\overline{I}_{SEICP-ROA}^{(p)-\alpha G}$ caused by the magnetic dipole response of the NP can be modeled as a sum of two effects: large magnetic field concentration at the position of the molecule and radiation enhancement of the magnetic dipole moment of the molecule. On the other hand, the amplification of $\overline{I}_{SEICP-ROA}^{(p)-\alpha A}$ by the magnetic dipole response of the NP should be understood as the symmetry breaking of the electric field from the incidental wave and the radiated field from the molecular electric dipole and electric quadrupole. The detailed discussions have been provided in Supporting Information 3. Furthermore, it's also worth to note that, although large $\bar{I}_{SEICP-ROA}^{(p)-\alpha G}$ caused by the magnetic response of the NP can be well predicted within the quasistatic approximation, the evaluation of $\bar{I}_{SEICP-ROA}^{(p)-\alpha A}$ relies on the electric fields scattered by the magnetic dipoles of the NP and the molecule, which are proportional to $1/r_0^2$, as can be seen from eq 6.

## III. NUMERICAL RESULTS AND DISCUSSION

**A. Enhanced SEROA of the Chiral Molecules by Various Kinds of Dipolar NPs.** In order to test the above analyses, we perform numerical calculations of the back-scattering ROA and Raman intensities of the chiral molecules ((+)-(R)-methyloxirane) enhanced by various kinds of dipolar NPs using the expressions derived above. The orientation averaging in eq 18 is performed numerically to take the retardation effect of the plane wave into account, rather than using the analytical formulas in ref 2. The parameters $\tilde{\alpha}$, $\tilde{G}'$ and $\tilde{A}$ for molecular response tensors are calculated using a development version of Gaussian at the excitation frequency of 510nm.[48,49] The geometry of methyloxirane is optimized at the B3LYP/aug-cc-pvTV computation level, which has been used by previous works.[50] Using the optimized ground state of the molecule, the response tensors were calculated using the same basis. Throughout the article, these molecular response tensors are evaluated with respect to the coordinate origin. This guarantees that the fields described in eqs 6 and 7 and molecular multipoles are evaluated at a common origin, which is essential

in achieving the origin-invariance of the calculated ROA signal.[51]

Note the chemical effects of the surface molecule are ignored in our model,[29,30] because the present work focuses on investigating various EM enhancement mechanisms of the SEROA rather than making quantitative comparison with experiments. In this paper, we limit our concentration on the nonadsorbed situation where the molecule may bind in any orientation with respect to the surface of the NP with the equal probability, and an average over the molecular rotation is performed throughout the text.

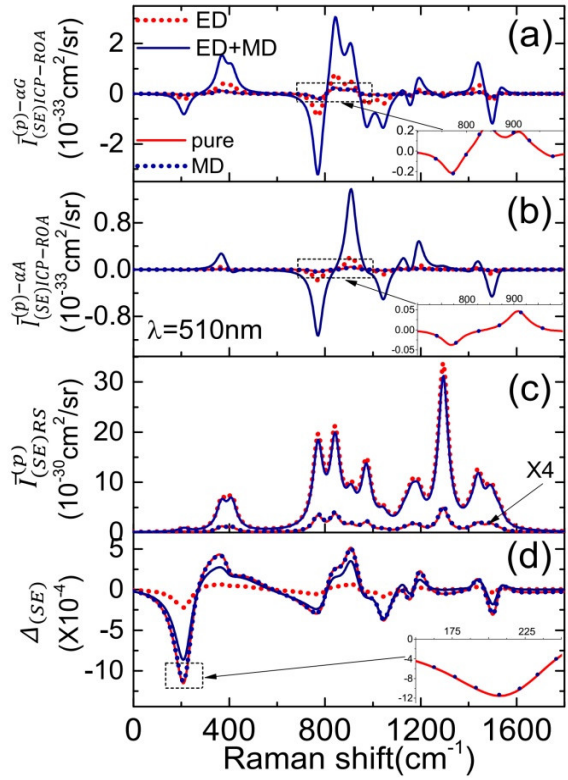

**Figure 2.** (a) $\tilde{\alpha}\tilde{G}$ components of ROA and SEROA spectra ($\bar{I}^{(p)-\alpha G}_{(SE)ICP-ROA}$). (b) $\tilde{\alpha}\tilde{A}$ components of ROA and SEROA spectra ($\bar{I}^{(p)-\alpha A}_{(SE)ICP-ROA}$). (c) Raman and SERS spectra ($\bar{I}^{(p)}_{(SE)RS}$). (d) Dimensionless (surface enhanced) circular intensity difference. Spectra are shown for (+)-(R)-methyloxirane on the NP with electric dipole response (ED), electric and magnetic dipole responses (ED+MD) or magnetic response (MD). Distances between the molecules and the center of the NP are set to be 11nm. Spectra for pure (R)-(+)-methyloxirane are shown as red lines for comparison. All the systems presented in figures are excited at the wavelength of 510nm. The spectra have been broadened by a Lorentzian with a full-width at half-maximum (fwhm) of $50 cm^{-1}$.

Figures 2a and b show calculated results for $\bar{I}_{(SE)ICP-ROA}^{(p)-\alpha G}$ and $\bar{I}_{(SE)ICP-ROA}^{(p)-\alpha A}$ as functions of Raman shift, respectively. A molecule is put 11nm ($r_0 = 11nm$) away from the center of the NP as illustrated in Figure 1. The red dotted lines represent the case of the NP with only electric dipole response (ED). Here the electric dipole polarizability of the NP is set to be equal with that of an Au NP which has a radius of 10nm, while the magnetic dipole polarizability is presumed to be zero. The electric dipole polarizability is calculated using the refraction indices of Au given in ref 52. The blue lines correspond to the case of the NP with both electric and magnetic dipole responses (ED+MD). Here the magnetic response of the NP is set to give a $S_m$ equal with $S_e$. Comparing them, we find that two components of the SEROA signal ($\bar{I}_{(SE)ICP-ROA}^{(p)-\alpha G}$ and $\bar{I}_{(SE)ICP-ROA}^{(p)-\alpha A}$) can be improved largely when magnetic dipole responses are considered. For example, the ROA signal $\bar{I}_{SEICP-ROA}^{(p)-\alpha G}$ is improved around 5 times, while $\bar{I}_{(SE)ICP-ROA}^{(p)-\alpha A}$ is enhanced by a factor of 9 at 842cm$^{-1}$. According to eq 19, the SEROA signal is proportional to the $S_m$. Thus, a strong ROA signal can be obtained by setting a larger $S_m$.

In fact, such a phenomenon is the result of combined action of ED and MD. If the NP has only magnetic dipole response and the electric dipole response is set to be zero, the results are described by blue dotted lines (MD). The red solid lines in Figures 2a and b show $\bar{I}_{ICP-ROA}^{(p)-\alpha G}$ and $\bar{I}_{ICP-ROA}^{(p)-\alpha A}$ for pure molecules in the free space. By comparison, we find that there is no improvement for two components of the ROA in such a case.

In addition to the ROA, one of key variable is the dimensionless circular intensity difference (CID) introduced by Barron and Buckingham of which the expression is[2-5]

$$\Delta_{SE} = \frac{\bar{I}_{SEICP-ROA}^{(p)}}{\bar{I}_{SERS}^{(p)}} \qquad (20)$$

where the parenting Raman signal $\bar{I}_{SERS}^{(p)}$ is given by

$$\bar{I}_{SERS}^{(p)} = K_p K_r \sum_{D=R;L} \left\langle \left| E_{rad,t,a}(D) \right|^2 \right\rangle_\Omega \qquad (21)$$

This variable characterizes the signal-to-background ratio of the measured system and determines the smallest ROA signal worth pursuing. If the value of $\Delta_{SE}$ is too small, the SEROA signal is very likely to be covered by the noise caused by the experiment setup.

Figures 2c and d show the $\bar{I}_{SERS}^{(p)}$ and the circular intensity differences $\Delta_{SE}$ for the systems discussed above as a function of Raman shift, respectively. Four kinds of case, ED (red dotted line), MD (blue dotted line), ED and MD (blue solid line), pure molecule (red solid line), have been considered. Comparing them, one finds $\bar{I}_{SERS}^{(p)}$ is nearly independent on the magnetic dipole response of the NP. Because the magnetic dipolar response of the NP can greatly enhance the ROA signal while keeping the parenting Raman signal unchanged, $\Delta_{SE}$ for the NP with magnetic responses is much larger than those possessing only electric dipole responses. As illustrated in Figure 2d, $\Delta_{SE}$ for the 'ED+MD' NP is around 5 times larger than that of the 'ED' NP. This is beneficial for the experiment realizations of the SEROA which are always plagued by the intrinsic low signal-to-background ratio.[11]

**B. SEROA by a Si Nano Sphere.** In the above parts, we concentrate our discussions on the SEROA by a deep subwavelength NP. It has been proved analytically that an enhanced ROA signal with a large circular intensity difference $\Delta_{SE}$ can be obtained by the NP with the magnetic dipole response. In this part, we study the SEROA by a Si NP, which have been shown both theoretically and experimentally to possess a strong magnetic dipolar response in the optical frequency.[37,38] Such a strong magnetic response is caused by the curl of intense displacement current induced by the external electromagnetic radiation, when the size of the particle is comparable to the effective wavelength in the dielectric material. Figure 3a describes the extinction spectrum (black lines), as well as the partial contributions from all kinds of multipoles for a Si NP with radius $R_s$=65nm. The blue line, red line and green line correspond to the electric dipole (ED), magnetic dipole (MD) and magnetic quadrupole (MQ) contributions of extinction, respectively. For comparison, corresponding results for an Au NP with $R_s$=65nm have also been plotted in Figure 3b. Refraction indices used in the calculation are taken from ref 53 for Si and ref 52 for Au.

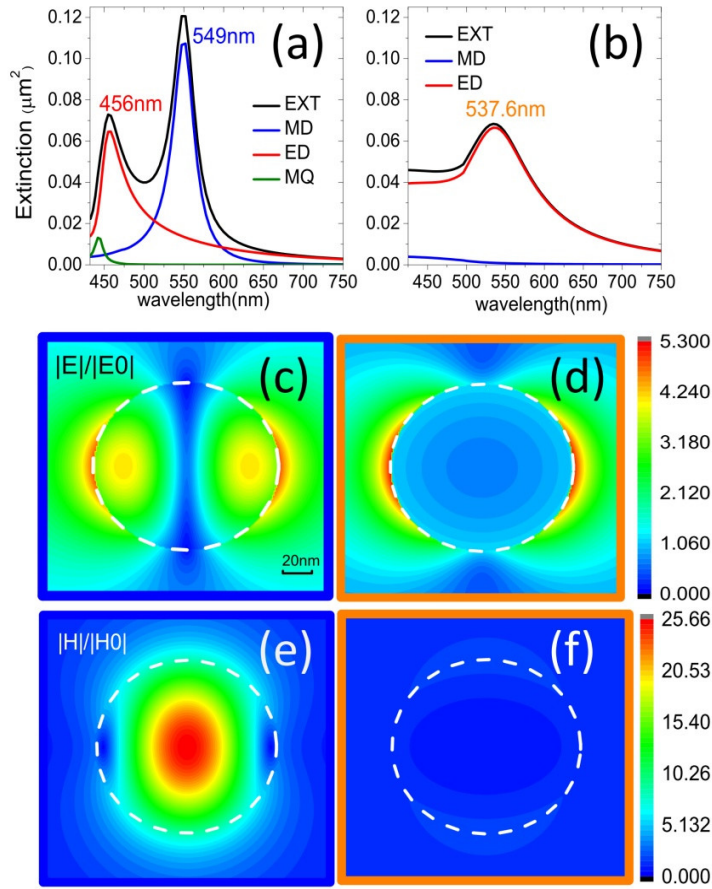

**Figure 3.** (a) Extinction cross section and its multipole decompositions for the Si NP with radius $R_s$=65nm. (b) The corresponding results for the Au NP. Spatial profiles of electric (c) and magnetic (e) field amplitudes for the Si NP with $R_s$=65nm at the wavelength of 549nm. The corresponding spatial profiles for the Au NP at the wavelength of 537.6nm are given by (d) and (f), respectively. The NPs are excited by X-polarized plane waves.

It can be seen clearly that a strong resonance peak in the extinction spectrum of the Si NP appears around the wavelength of 549nm, which is in contrast to the case for the Au NP. Such a peak is dominated by the magnetic dipole resonance of the Si NP, which makes it a good candidate for the SEROA according to the discussions in the part II. In order to disclose the phenomenon further, the near-field amplitude maps for Si and Au NPs with $R_s$=65nm excited by x-polarized plane waves propagating in the z direction are plotted in Figures 3c-f. Figures 3c and e exhibit distributions of electric and magnetic fields for the Si NP with $R_s$=65nm at the wavelength of 549nm. The electric field profile (Figure 3c) shows two-lobe distribution which is a typical character of the electric dipole mode. Besides, one can observe two circular electric field concentration regions inside the sphere, which signifies that a magnetic dipole along the y direction is induced. Due to the existence of the magnetic dipole resonance, the magnetic field is strongly concentrated inside and near the NP as shown in Figure 3e. According to the theory described in part II, the concentration of both electric and magnetic fields inside the Si NP may lead to a larger $\overline{I}_{SEICP-ROA}^{(p)-\alpha G}$. In contrast, the magnetic dipole excitation for the Au NP at the resonance wavelength of 537.6nm is very weak and the concentration of the magnetic field inside the Au NP is almost zero as shown in Figure 3f, although the electric dipole resonance is very strong as shown in Figure 3d.

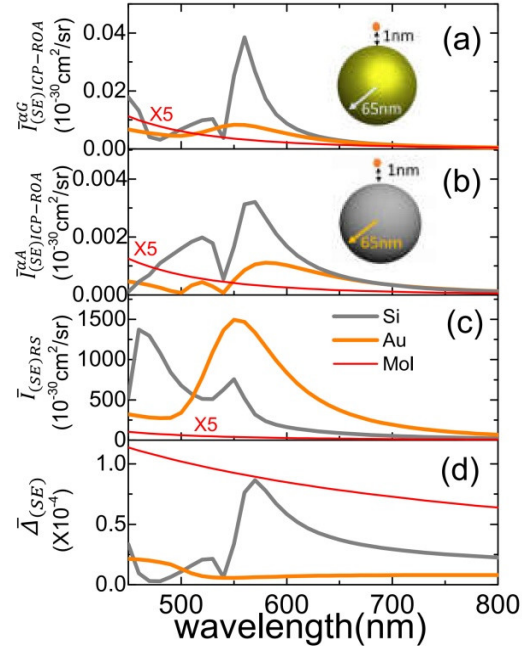

**Figure 4.** Spectra for (R)-CHFClBr enhanced by a Si sphere with $R_s$=65nm (grey lines) and an Au sphere with $R_s$=65nm (orange lines). Distances between the molecules and the surface of the NP are set to be 1nm. (a) Mode averaged $\tilde{\alpha}\tilde{G}$ components of ROA and SEROA spectra ($\bar{I}^{\alpha G}_{(SE)ICP-ROA}$). (b) Mode averaged $\tilde{\alpha}\tilde{A}$ components of ROA and SEROA spectra ($\bar{I}^{\alpha A}_{(SE)ICP-ROA}$). (c) Mode averaged surface enhanced or unenhanced Raman scattering intensities ($\bar{I}_{(SE)RS}$). (d) Mode averaged circular intensity difference $\bar{\Delta}_{(SE)}$. The red lines in (a)-(c) represent the results without the NP, which is multiplied by an amplification factor 5.

In Figures 4a and b, we present the comparison of the calculated results of mode averaged $\tilde{\alpha}\tilde{G}$ and $\tilde{\alpha}\tilde{A}$ components of the ROA for a chiral molecule (R)-CHFClBr near Si and Au NPs with $R_s$=65nm. Distance between the molecule and the center of the NP is set as 66nm. The mode averaged $\tilde{\alpha}\tilde{G}$ and $\tilde{\alpha}\tilde{A}$ components of (surface enhanced) ROA are defined as

$$\overline{I}_{(SE)ICP-ROA}^{\alpha G} = \sum_p \left| \overline{I}_{(SE)ICP-ROA}^{(p)-\alpha G} \right| \tag{22}$$

$$\overline{I}_{(SE)ICP-ROA}^{\alpha A} = \sum_p \left| \overline{I}_{(SE)ICP-ROA}^{(p)-\alpha A} \right| \tag{23}$$

where the contributions of all modes are taken into account. Then

$$\overline{I}_{(SE)ICP-ROA} = \sum_p \left| \overline{I}_{(SE)ICP-ROA}^{(p)} \right| \tag{24}$$

$$\overline{I}_{(SE)RS} = \sum_p \overline{I}_{(SE)RS}^{(p)} \tag{25}$$

Here $\overline{I}_{(SE)ICP-ROA}$ and $\overline{I}_{(SE)RS}$ represent mode averaged SEROA and SERS, respectively. We can also define the difference of the mode averaged circular intensity as

$$\overline{\Delta}_{(SE)} = \frac{\overline{I}_{(SE)ICP-ROA}}{\overline{I}_{(SE)RS}} \tag{26}$$

Similar to $\Delta_{SE}$, $\overline{\Delta}_{(SE)}$ can also be used to characterize the signal-to-background ratio in the ROA experiment.

The results have been obtained by using an extended Mie scattering theory.[26,27] The parameters of Si and Au NPs are taken identical with those in Figure 3. The wavelength dependent molecular response tensors $\tilde{\alpha}$, $\tilde{G}'$ and $\tilde{A}$ are obtained from B3YLYP/6-31G** calculations at various excitation frequencies (from 450nm to 800nm).[49] The grey and orange lines correspond to the results for the Si and Au NPs, respectively. Signals for pure molecule are also presented as red lines for comparison. A strong resonance improvement of both $\overline{I}_{(SE)ICP-ROA}^{\alpha G}$ and $\overline{I}_{(SE)ICP-ROA}^{\alpha A}$ appears around the magnetic dipole resonance wavelength of 549nm for the Si NP. In such a case, $\overline{I}_{SEICP-ROA}^{\alpha G}$ for the Si NP is around 5 times larger than that for the Au NP, while the value of $\overline{I}_{SEICP-ROA}^{\alpha A}$ is about 3 times larger.

The corresponding results for $\overline{I}_{(SE)RS}$ and $\overline{\Delta}_{(SE)}$ are shown in Figures 4c and d, respectively. The enhancement factors for the SERS are on the order of $10^2$ which are identical with the experimental results.[54] In contrast to the deep subwavelength NP discussed in Figure 2, both electric and magnetic dipole response peaks can be observed in the spectrum of the Si NP. However, the SERS enhancement factor for the Si NP is still about 2 times smaller than that for the Au NP at the magnetic dipole response peak (549nm). Thus, it can be expected the circular intensity difference for the Si NP will be much larger than that for the Au NP. From Figure 4d, one can find the $\overline{\Delta}_{(SE)}$ for the Si NP is close to that of pure molecule at the wavelength of the magnetic dipole resonance and is 10 times larger than that of the Au NP.

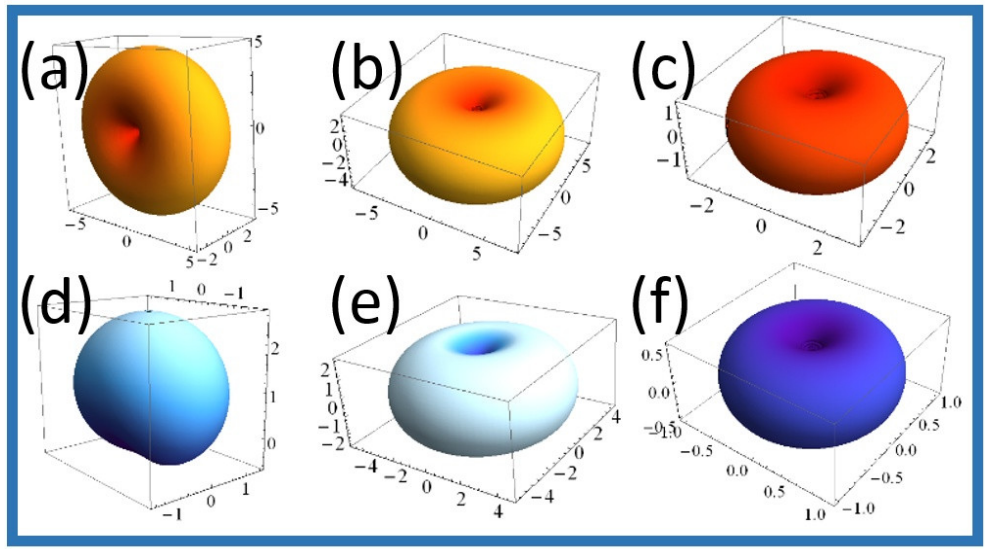

**Figure 5.** The corresponding far field radiation patterns $\lim_{r\to\infty}\left|rE^{(ED/MD)}(\boldsymbol{r})\right|^2$ for electric (a-c) and magnetic (b-d) dipoles. (a), (b), (d) and (e) represent the results for the Si sphere with $R_s$=65nm when the dipole is put away 1nm from the surface of the sphere. (c) and (f) are the corresponding case for the dipole without the NP. (a) and (d) represent directions of the dipole perpendicular to the surface of the Si NP, (b) and (e) correspond to the case parallel with the surface. The dipoles radiate at the wavelength of 549nm.

The Si NP not only causes the magnetic field from the incidental wave to strongly concentrate at the position of the molecule, but also has a strong impact on the molecular electric and magnetic dipole radiations. According to the discussions above, large $\tilde{\alpha}\tilde{G}$ components of the SEROA may be caused by enhancements of both electric and magnetic dipole radiations of the molecule. Figure 5 shows angular distributions of the scattering field intensity ($\lim_{r\to\infty}\left|rE^{(ED/MD)}(\boldsymbol{r})\right|^2$) for electric and magnetic dipoles. Figure 5a is the case when an electric dipole is put 66nm away from the center of a 65nm Si NP. The direction of the dipole is set to be perpendicular with the symmetry axis of the NP. Figure 5b is same with Figure 5a, but the dipole is parallel with the axis. The corresponding results for the electric dipole radiation in the absence of NP is presented in Figure 5c. Figures 5d-f are arranged analogous to Figures 5a-c, but the electric dipoles are substituted by magnetic dipoles. The dipoles are radiated at the wavelength of 549nm.

Comparing Figures 5a-b with Figure 5c, one can find the radiation intensity of the electric dipole is enhanced by a factor of ~2. On the other hand, from Figure 5d-f, radiation intensities of the magnetic dipole are amplified by a factor close to 2 for the perpendicular case and 4 for the parallel case.

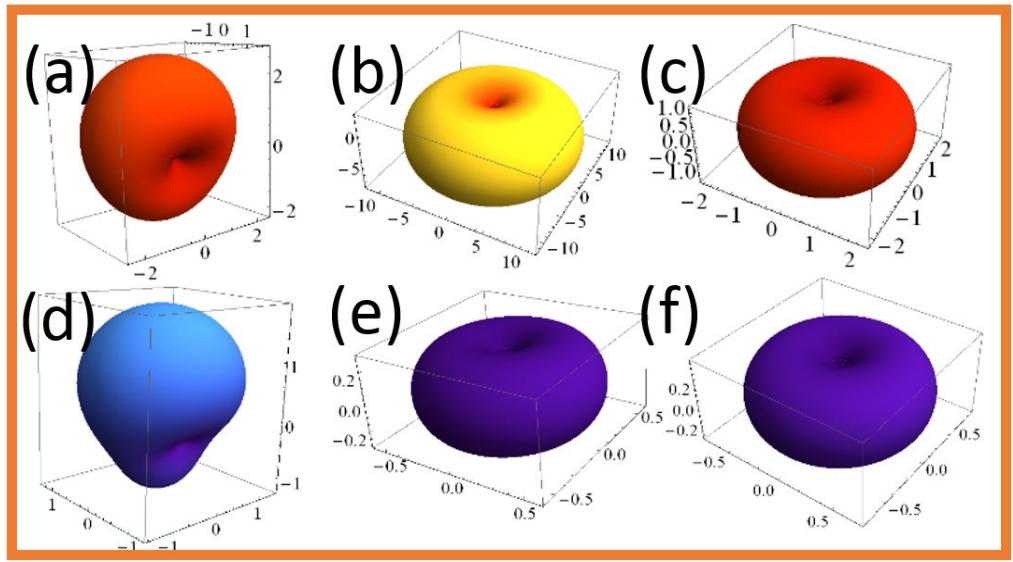

**Figure 6.** The corresponding results with Figure 5 but the Si NPs are substituted by Au NPs, and the dipoles radiate at the wavelength of 537.6nm.

In order to make comparison, the radiation patterns for dipoles being put near the surface of the Au NP are also plotted in Figure 6. Figure 6 is arranged the same to Figure 5, but the Si NP is substituted by the corresponding Au NP and the wavelength of the dipole radiation is 537.6nm. As can be found from Figures 6b and c, the radiation from the electric dipole parallel with the symmetry axis (Figure 6b) are significantly enhanced by the Au NP. The enhancement factor can be as large as 10. Comparing Figures 6d and e with Figure 6f, the radiation intensity is only enhanced when the magnetic dipole is perpendicular with the symmetry axis of the NP. When the magnetic dipole is set to be parallel with the norm of the sphere surface, the effect of the NP can even be ignored. Such a nonuniform enhancement of electric and magnetic dipole radiations for the molecule explains why the Au NP is inferior in enhancing the ROA.

In addition, the improvement of $\bar{I}_{(SE)ICP-ROA}^{\alpha A}$ as shown in Figure 4b originates from the symmetry breaking caused by the magnetic dipole response of the Si NP, which can not be explained as larger electric field gradient or enhanced molecular quadrupole radiation (the detailed numerical verification is given in Supporting Information 5).

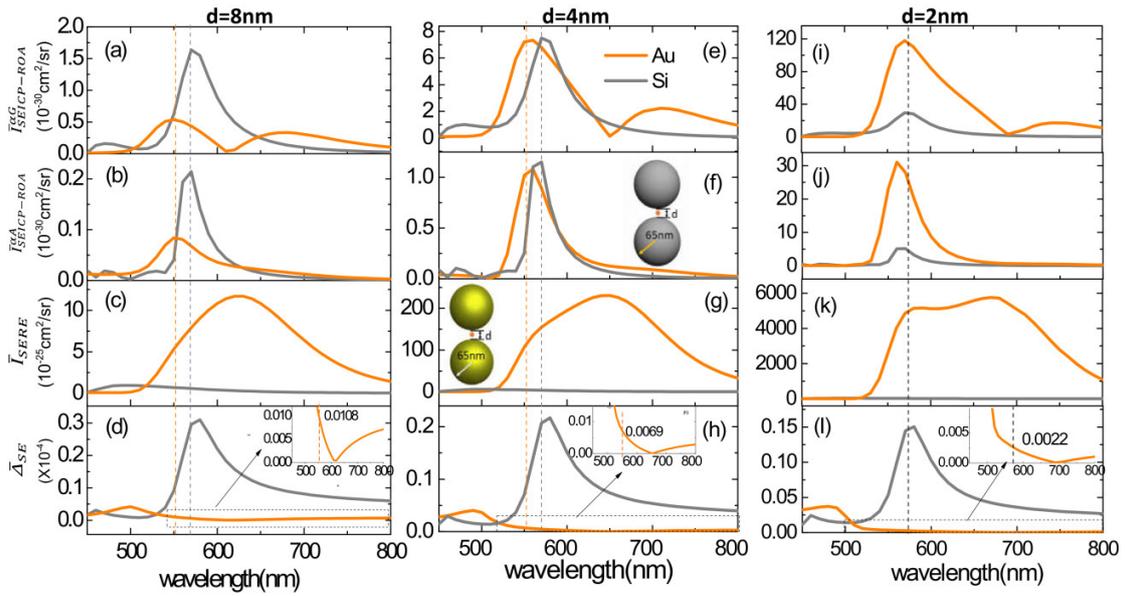

**Figure 7.** (a), (e) and (i) describe $\alpha G$ contributions to the SEROA ($\bar{I}_{SEICP-ROA}^{\alpha G}$) at d=8nm, 4nm and 2nm, respectively; The corresponding $\alpha A$ contributions to the SEROA ($\bar{I}_{SEICP-ROA}^{\alpha A}$) are given in (b), (f) and (j). (c), (g) and (k) correspond to the surface enhanced Raman scattering intensity ($\bar{I}_{SERS}$); The corresponding surface enhanced mode averaged circular intensity differences ($\bar{\Delta}_{SE}$) are described by (d), (h) and (l). Grey lines correspond

to a Si dimer with NPs of $R_s$=65nm. Orange lines to the corresponding Au dimer. The other parameters are identical with those in Figure 4.

**C. SEROA by the Hotspots in the Si Dimer.** In this part, we discuss the SEROA of a chiral molecule ((R)-CHFClBr) when it is put in the hotspots of a Si NP dimer with various separation distances.[55-58] The grey solid lines in Figure 7 present the calculated results for these cases. For comparison, the calculated results for the Au NP dimer are described by the orange lines. Here the parameters and sizes of Si and Au spheres are taken identical with those in Figure 4. Figures 7a-d correspond to the case with the separation distance between two spheres d=8nm, Figures 7e-h to that with d=4nm, and Figures 7i-l to that with d=2nm. Comparing Figures 7a and b with Figures 4a and b, one can find both $\bar{I}_{SEICP-ROA}^{(p)-\alpha G}$ and $\bar{I}_{SEICP-ROA}^{(p)-\alpha A}$ are significantly enhanced in the present case. For example, the value of $\bar{I}_{SEICP-ROA}^{(p)-\alpha G}$ for the Si dimer at the peak is improved by a factor around 40 compared with that of the single Si NP. With the decrease of the separation distance, both $\bar{I}_{SEICP-ROA}^{(p)-\alpha G}$ and $\bar{I}_{(SE)ICP-ROA}^{(p)-\alpha A}$ increase rapidly. For instance, the peak value of $\bar{I}_{SEICP-ROA}^{(p)-\alpha G}$ for the Si dimer with d=4nm is 4.5 times larger than that with d=8nm.

In fact, the increases of $\bar{I}_{SEICP-ROA}^{(p)-\alpha G}$ and $\bar{I}_{(SE)ICP-ROA}^{(p)-\alpha A}$ for the Au NP dimer are faster than those of the Si NP dimer with the decrease of the separation distance due to the strong field focusing effect in the hotspots. When d=8nm, the peak values of $\bar{I}_{SEICP-ROA}^{(p)-\alpha G}$ and $\bar{I}_{(SE)ICP-ROA}^{(p)-\alpha A}$ for the Au dimer are smaller than those for the Si dimer. However, these values for the Au dimer are much larger than those for the Si dimer at d=2nm. At the same time, from Figure 7c, g and k, we find that the SERS for the Au dimer also becomes larger with the decrease of the separation distance, which results in that the mode averaged circular intensity difference ($\bar{\Delta}_{SE}$) for the Au dimeris always smaller than that of the Si dimer.

From Figure 7(d) and (h), it can be seen that the values of $\bar{\Delta}_{SE}$ at peaks (570nm) of $\bar{I}_{SEICP-ROA}^{(p)-\alpha G}$ and $\bar{I}_{(SE)ICP-ROA}^{(p)-\alpha A}$ for the Si dimers are about 30 times larger than those of the Au dimers (peaks at the wavelength of 550nm). With the decreasing of the separation distance, the phenomenon becomes more remarkable. For example, the $\bar{\Delta}_{SE}$ for the Si dimer is 60 times larger than that of Au dimer at the wavelength of 570nm (peak of SEROA for both Au and Si dimer) as shown in Figure 7l. This means that the signal-to-background ratio for the Si structure is much larger than that of the Au dimer. Thus, it is very advantageous to use the present Si structure to observe the SEROA signal. Recently, some clusters of gold nanospheres, i.e. the dimers/trimers/tetramers, were fabricated successfully by using cysteine chiral molecules as linkers at the hotspots.[59] We expect the present Si clusters with molecules can be fabricated and the phenomena can be observed experimentally in the future.

**Conclusion**

In conclusion, we have demonstrated analytically and numerically that the SEROA signal of chiral molecules can be strongly enhanced by NPs with the strong magnetic dipole response. Furthermore, the ROA to Raman ratio of the SEROA enhanced by NPs with strong magnetic response are much larger than that of plasmonic NPs. Different mechanics are assigned to the enhancements of $\tilde{\alpha}\tilde{G}$ and $\tilde{\alpha}\tilde{A}$ parts of SEROA signals. The former is enhanced due to the larger magnetic field at the position of molecule and the radiation enhancement of the magnetic dipole while the latter is amplified by the symmetry breaking introduced by the magnetic dipole response of the NP. We have taken Si and Au NPs as examples and found that the SEROA signal of the Si NP can be ~5 times larger than that of the Au NP, while CID can be improved by a factor of ~10. At last, we have studied the dimer systems formed by Au and Si NPs. We have calculated the SEROA for cases when the chiral molecules are put at the hotspots of the dimers. Although the SEROA signal of the Si dimer is close to that of the Au dimer, the ROA to Raman ratio is about 60 times larger. It is very beneficial for observing the SEROA signal to use the present Si structures.

**AUTHOR INFORMATION**

**Corresponding Author**

* E-mail: zhangxd@bit.edu.cn

**Notes**

The authors declare no competing financial interest.

**Acknowledgment**

This work was supported by the National Natural Science Foundation of China (Grant No. 11274042, 61421001and 11174033)

**Supporting Information**

Details of analytical derivations for theory and method are given. This material is available free of charge via the Internet at http://pubs.acs.org.

**References**

(1) *Circular Dichroism and the Conformational Analysis of Biomolecules*; Fasman, G. D., Ed.; Plenum: New York, 1996.

(2) Polavarapu, P. L., *Vibrational spectra: principles and applications with emphasis on optical activity*. Elsevier: Amsterdam, The Netherlands, 1998.

(3) Barron, L. D. *Molecular Light Scattering and Optical Activity*, 2nd ed.; Cambridge University Press: 2004.

(4) Nafie, L. A. *Vibrational Optical Activity: Principles and Applications*; John Wiley & Sons, Ltd.: New York, 2011.

(5) Barron, L.; Buckingham, A. Rayleigh and Raman Scattering from Optically Active Molecules. *Mol. Phys.* **1971**, *20*, 1111−1119.

(6) Hug, W. Visualizing Raman and Raman optical activity generation in polyatomic molecules. *Chem. Phys.* **2001**, *264*, 53-69.

(7) Parchaňský, V.; Kapitán, J.; Bouř, P. Inspecting Chiral Molecules by Raman Optical Activity Spectroscopy. *RSC Adv.* **2014**, *4*, 57125−57136.

(8) McColl, L. H.; Blanch, E. W.; Gill, A. C.; Rhie, A. G. O.; Ritchie, M. A.; Hecht, L.;

Nielsen, K.; Barron, L. D. A New Perspective on β-Sheet Structures Using Vibrational Raman Optical Activity: From Poly(l-lysine) to the Prion Protein. *J. Am. Chem. Soc.* **2003**, *125*, 10019–10026.

(9) Kapitan, J.; Baumruk, V.; Kopecký, V.; Bouř, P. Conformational flexibility of L-alanine zwitterion determines shapes of Raman and Raman optical activity spectral bands. *J. Phys. Chem. A* **2006**, *110*, 4689-4696.

(10) Mutter, S. T.; Zielinski, F.; Popelier, P. L.; Blanch, E. W. Calculation of Raman optical activity spectra for vibrational analysis. *Analyst* **2015**, *140*, 2944-2956.

(11) Abdali, S.; Blanch, E. W. Surface Enhanced Raman Optical Activity (SEROA). *Chem. Soc. Rev.* **2008**, *37*, 980−992.

(12) Johannessen, C.; White, P. C.; Abdali, S. Resonance Raman Optical Activity and Surface Enhanced Resonance Raman Optical Activity Analysis of Cytochrome c. *J. Phys. Chem. A* **2007**, *111*, 7771−7776.

(13) Abdali, S.; Johannessen, C.; Nygaard, J.; Nørbygaard, T. Resonance Surface Enhanced Raman Optical Activity of Myoglobin as a Result of Optimized Resonance Surface Enhanced Raman Scattering Conditions. *J. Phys.: Condens. Matter* **2007**, *19*, 285205.

(14) Osinska, K.; Pecul, M.; Kudelski, A. Circularly Polarized Component in Surface-enhanced Raman Spectra. *Chem. Phys. Lett*. **2010**, *10*, 86−90.

(15) Brinson, B. Nonresonant Surface Enhanced Raman Optical Activity, Ph.D. thesis, Rice University, Houston, TX, 2009.

(16) Pour, S. O.; Bell, S. E. J.; Blanch, E. W. Use of a hydrogel polymer for reproducible

surface enhanced Raman optical activity (SEROA). *Chem. Commun.* **2011**, *47*, 4754-4756.

(17) Kneipp, H.; Kneipp, J.; Kneipp, K. Surface-enhanced Raman optical activity on adenine in silver colloidal solution. *Anal. Chem.* **2006**, *78*, 1363-1366.

(18) Sun, M.; Zhang, Z.; Wang, P.; Li, Q.; Ma, F.; Xu, H. Remotely Excited Raman Optical Activity Using Chiral Plasmon Propagation in Ag Nanowires. *Light: Sci. Appl.* **2013**, *2*, e112.

(19) Abdali, S. Observation of SERS Effect in Raman Optical Activity, a New Tool for Chiral Vibrational Spectroscopy. *J. Raman Spectrosc.* **2006**, *37*, 1341−1345.

(20) Etchegoin, P. G.; Galloway, C.; Le Ru, E. C. Polarization-dependent effects in surface-enhanced Raman scattering(SERS). *Phys. Chem. Chem.Phys*. **2006**, *8*, 2624.

(21) Efrima, S. Raman Optical Activity of Molecules Adsorbed on Metal Surfaces: Theory. *J. Chem. Phys.* **1985**, *83*, 1356−1362.

(22) Efrima, S. The Effect of Large Electric Field Gradients on the Raman Optical Activity of Molecules Adsorbed on Metal Surfaces. *Chem. Phys. Lett.* **1983**, *102*, 79−82.

(23) Janesko, B. G.; Scuseria, G. E. Surface Enhanced Raman Optical Activity of Molecules on Orientationally Averaged Substrates: Theory of Electromagnetic Effects. *J. Chem. Phys.* **2006**, *125*, 124704.

(24) Bouř, P. Matrix Formulation of the Surface-enhanced Raman Optical Activity Theory. *J. Chem. Phys.* **2007**, *126*, 136101.

(25) Novák, V.; Šebestík, J.; Bouř, P. Theoretical Modeling of the Surface-Enhanced Raman Optical Activity. *J. Chem. Theory Comput.* **2012**, *8*, 1714−1720.

(26) Acevedo, R.; Lombardini, R.; Halas, N. J.; Johnson, B. R. Plasmonic Enhancement of Raman Optical Activity in Molecules near Metal Nanoshells. *J. Phys. Chem. A* **2009**, *113*, 13173−13183.

(27) Lombardini, R.; Acevedo, R.; Halas, N. J.; Johnson, B. R. Plasmonic Enhancement of Raman Optical Activity in Molecules near Metal Nanoshells: Theoretical Comparison of Circular Polarization Methods. *J. Phys. Chem. C* **2010**, *114*, 7390−7400.

(28) Chulhai, D.; Jensen, L. Simulating Surface-Enhanced Raman Optical Activity Using Atomistic Electrodynamics-Quantum Mechanical Models. *J. Phys. Chem. A* **2014**, *118*, 9069−9079.

(29) Janesko, B. G.; Scuseria, G. E. Molecule-Surface Orientational Averaging in Surface Enhanced Raman Optical Activity Spectroscopy. *J. Phys. Chem. C* **2009**, *113*, 9445−9449.

(30) Jensen, L. Surface-Enhanced Vibrational Raman Optical Activity: A Time-Dependent Density Functional Theory Approach. *J. Phys. Chem. A* **2009**, *113*, 4437−4444.

(31) Le Ru, E. C.; Etchegoin, P. G. Principles of Surface-Enhanced Raman Spectroscopy and Related Plasmonic Effects; Elsevier: Amsterdam, 2009.

(32) Nie, S.; Emory, S. R. Probing Single Molecule and Single Nanoparticles by Surface-Enhanced Raman Scattering. *Science* **1997**, *275*, 1102−1106.

(33) Wei, F.; Zhang, D.; Halas, N. J.; Hartgerink, J. D. Aromatic amino acids providing characteristic motifs in the Raman and SERS spectroscopy of peptides. *J. Phys. Chem. B* **2008**, *112*, 9158-9164.

(34) Verre, R.; Yang, Z. J.; Shegai, T.; Kall, M. Optical magnetism and plasmonic Fano

resonances in metal–insulator–metal oligomers. *Nano Lett.* **2015**, *15*, 1952-1958.

(35) Sheikholeslami, S. N.; Garcia-Etxarri, A.; Dionne, J. A. Controlling the Interplay of Electric and Magnetic Modes via Fano-like Plasmon Resonances. *Nano Lett*. **2011**, *11*, 3927−3934.

(36) Cortie, M.; Ford, M. A plasmon-induced current loop in gold semi-shells. *Nanotechnology* **2007**, *18*, 235704.

(37) García-Etxarri, A.; Gómez -Medina, R.; Froufe-Pérez, L. S.; López, C.; Chantada, L.; Scheffold, F.; Aizpurua, J.; Nieto-Vesperinas, M.; Sáenz, J. J. *Opt. Express* **2011**, *19*, 4815−4826.

(38) Evlyukhin, A. B.; Novikov, S. M.; Zywietz, U.; Eriksen, R. L.; Reinhardt, C.; Bozhevolnyi, S. I.; Chichkov, B. N. Demonstration of magnetic dipole resonances of dielectric nanospheres in the visible region. *Nano Lett.* **2012**, *12*, 3749-3755.

(39) Alizadeh, M. H.; Reinhard, B. M. Plasmonically enhanced chiral optical fields and forces in achiral split ring resonators. *ACS Photonics* **2015**, *2*, 361-368.

(40) García-Etxarri, A.; Dionne, J. A. Surface-enhanced circular dichroism spectroscopy mediated by nonchiral nanoantennas. *Phys. Rev. B* **2013**, *87*, 235409.

(41) Finazzi, M.; Biagioni, P.; Celebrano, M.; Duò, L. Quasistatic limit for plasmon-enhanced optical chirality. *Phys. Rev. B* **2015**, *91*, 195427.

(42) Born, M.; Huang, K. Dynamical Theory of Crystal Lattices; Clarendon Press: Oxford, U.K., 1954.

(43) Doicu, A.; Wriedt, T.; Eremin, Y. A. Light Scattering by Systems of Particles;

Springer-Verlag: Berlin, 2006.

(44) Evlyukhin, A. B.; Reinhardt, C.; Zywietz, U.; Chichkov, B. N. Collective resonances in metal nanoparticle arrays with dipole-quadrupole interactions. *Phys. Rev. B* **2012**, *85*, 245411.

(45) Hecht, L.; Barron, L. D. Rayleigh and Raman optical activity from chiral surfaces and interfaces. *J. Mol. Struct.* **1995**, *348*, 217.

(46) Monticone, F.; Alù, A. The quest for optical magnetism: from split-ring resonators to plasmonic nanoparticles and nanoclusters. *J. Mater. Chem. C* **2014**, *2*, 9059-9072.

(47) Qian, Z.; Hastings, S. P.; Li, C.; Edward, B.; McGinn, C. K.; Engheta, N.; Fakhraai, Z.; Park, S.-J. Raspberry-like Metamolecules Exhibiting Strong Magnetic Resonances. *ACS Nano* **2015**, *9,* 1263– 1270.

(48) Cheeseman, J. R.; Frisch, M. J. Basis Set Dependence of Vibrational Raman and Raman Optical Activity Intensities. *J. Chem. Theory Comput.* **2011**, *7*, 3323−3334.

(49) Frisch, M. J.; Trucks, G. W.; Schlegel, H. B.; Scuseria, G. E.; Robb, M. A.; Cheeseman, J. R.; Scalmani, G.; Barone, V.; Mennucci, B.; Petersson, G. A.; etal. Gaussian09; Gaussian, Inc.: Wallingford, CT, 2009.

(50) Šebestík, J.; Bouř P. Raman optical activity of methyloxirane gas and liquid. *J. Phys. Chem. Lett.* **2011**, *2*, 498–502.

(51) Chulhai, D. V.; Jensen, L. Determining molecular orientation with surface-enhanced raman scattering using inhomogenous electric fields. *J. Phys. Chem. C* **2013**, *117*, 19622−19631.

(52) Johnson, P. B.; Christy, R. W. Optical Constants of the Noble Metals. *Phys. Rev. B: Solid State* **1972**, *6*, 4370−4379.

(53) Vuye, G.; Fisson, S.; Van, V. N.; Wang, Y.; Rivory, J.; Abeles, F. Temperature dependence of the dielectric function of silicon using in situ spectroscopic ellipsometry. *Thin Solid Films* **1993**, *233*, 166-170.

(54) Dmitriev, P. A., Baranov, D. G., Milichko, V. A., Makarov, S. V., Mukhin, I. S., Samusev, A. K., Krasnok, A. E., Belov, P. A., Kivshar, Y. S. Resonant Raman Scattering from Silicon Nanoparticles Enhanced by Magnetic Response. **2016**, *arXiv*:1601.03757.

(55) Caldarola, M.; Albella, P.; Cortes, E.; Rahmani, M.; Roschuk, T.; Grinblat, G.; Oulton, R. F.; Bragas, A. V.; Maier, S. A. Non-plasmonic Nanoantennas for Surface Enhanced Spectroscopies with Ultra-low Heat Conversion *Nat. Commun*. **2015**, *6*, 7915− 7915.

(56) Bakker, R. M.; Permyakov, D.; Yu, Y. F.; Markovich, D.; Paniagua-Domínguez, R.; Gonzaga, L.; Samusev, A.; Kivshar, Y.; Luk'yanchuk, B.; Kuznetsov, A. I. Magnetic and Electric Hotspots with Silicon Nanodimers *Nano Lett*. **2015**, *15*, 2137− 2142.

(57) Albella, P.; Poyli, M. A.; Schmidt, M. K.; Maier, S. A.; Moreno, F.; Saenz, J. J.; Aizpurua, J. Low-loss electric and magnetic fieldenhanced spectroscopy with subwavelength silicon dimers. *J. Phys. Chem. C* **2013**, *117*, 13573−13584.

(58) Yan, J.; Liu, P.; Lin, Z.; Wang, H.; Chen, H.; Wang, C.; Yang, G. Directional Fano resonance in a silicon nanosphere dimer. *ACS Nano* **2015**, *9*, 2968−2980.

(59) Wang, R. Y.; Wang, P.; Liu, Y.; Zhao, W.; Zhai, D.; Hong, X.; Ji, Y.; Wu, X.; Wang, F.; Zhang, D.; etal. Experimental observation of giant chiroptical amplification of small chiral molecules by gold nanosphere clusters. *J. Phys. Chem. C* **2014**, *118*, 9690−9695.

**The Table of Contents graphic:**

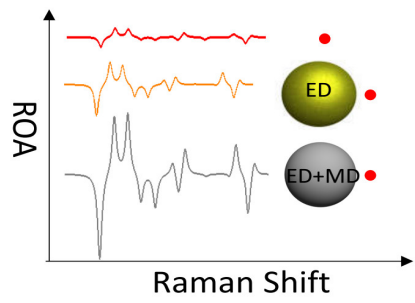